\documentclass[epj,nopacs]{svjour}
\pdfoutput=1
\usepackage[utf8]{inputenc}
\usepackage[colorlinks=true,urlcolor=blue,anchorcolor=blue,citecolor=blue,filecolor=blue,linkcolor=blue,menucolor=blue,pagecolor=blue,linktocpage=true,pdfa=true,unicode=true,bookmarksopen=true]{hyperref}

\usepackage{mathtools}
\usepackage{lmodern}
\usepackage[T1]{fontenc}
\usepackage{bbm}
\DeclareMathAlphabet{\mathbfi}{OML}{cmm}{b}{it}

\let\originalleft\left
\let\originalright\right
\renewcommand{\left}{\mathopen{}\mathclose\bgroup\originalleft}
\renewcommand{\right}{\aftergroup\egroup\originalright}
\makeatletter
\newcommand{\vast}{\bBigg@{4}}
\newcommand{\Vast}{\bBigg@{5}}
\makeatother

\makeatletter
\newenvironment{equations}[1][]{\subequations\ifx\relax#1\relax\else\label{#1}\fi\align\ignorespaces}{\endalign\ignorespacesafterend\endsubequations}
\def\@spliteq#1{\begin{equation}\begin{split}#1\end{split}\end{equation}}
\def\splitequation{\collect@body\@spliteq}

\makeatother

\renewcommand{\vec}[1]{{\ifnum9<1#1\mathbf{#1}\else\ifcat\noexpand#1\relax\boldsymbol{#1}\else\mathbfi{#1}\fi\fi}}

\let\oldre\Re
\let\oldim\Im
\renewcommand{\Re}{\oldre\mathfrak{e}\,}
\renewcommand{\Im}{\oldim\mathfrak{m}\,}
\newcommand{\total}{\mathop{}\!\mathrm{d}}

\newcommand{\abs}[1]{{\left\lvert{#1}\right\rvert}}

\newcommand{\eqend}[1]{\,#1}
\newcommand{\bigo}[1]{\mathcal{O}\left({#1}\right)}

\newcommand{\overbullet}[1]{\overset{\bullet}{#1}\vphantom{#1}}
\newcommand{\overring}[1]{\overset{\circ}{#1}\vphantom{#1}}

\allowdisplaybreaks

\makeatletter
\gdef\@fpheader{\ }
\makeatother

\begin{document}

\title{Kerr--Schild metrics in teleparallel gravity}

\author{Markus B. Fr{\"o}b}
\institute{Institut f{\"u}r Theoretische Physik, Universit{\"a}t Leipzig, Br{\"u}derstra{\ss}e 16, 04103 Leipzig, Germany\\\email{\href{mailto:mfroeb@itp.uni-leipzig.de}{mfroeb@itp.uni-leipzig.de}}}

\abstract{We show that the Kerr--Schild ansatz can be extended from the metric to the tetrad, and then to teleparallel gravity where curvature vanishes but torsion does not. We derive the equations of motion for the Kerr--Schild null vector, and describe the solution for a rotating black hole in this framework. It is shown that the solution depends on the chosen tetrad in a non-trivial way if the spin connection is fixed to be the one of the flat background spacetime. We show furthermore that any Kerr--Schild solution with a flat background is also a solution of $f(\mathcal{T})$ gravity.}

\date{Received: 11 March 2021 / Accepted: 16 August 2021 / Published: 24 August 2021}


\maketitle

\section{Introduction}
\label{sec_intro}

The Kerr--Schild ansatz~\cite{trautman1962,kerr1963,kerrschild2009,debneyetal1969}, in which the full metric $g_{\mu\nu} = \bar{g}_{\mu\nu} + 2 F k_\mu k_\nu$ is expressed as the sum of a background metric $\bar{g}_{\mu\nu}$ and the tensor product of a null vector $k_\mu$ (rescaled by a scalar function $F$),  has been very successful in determining exact solutions of Einstein gravity. This follows from the fact that linear perturbation theory with this ansatz is exact~\cite{gursesgursey1975}: the inverse metric is given by $g^{\mu\nu} = \bar{g}^{\mu\nu} - 2 F k^\mu k^\nu$, and the Ricci tensors of the full and background metric are related by
\begin{splitequation}
\label{kerrschild_ricci}
R^\mu{}_\nu &= \bar{R}^\mu{}_\nu - 2 F k^\mu k^\rho \bar{R}_{\rho\nu} \\
&+ \bar{\nabla}_\rho \left[ \bar{\nabla}^\mu \left( F k^\rho k_\nu \right) + \bar{\nabla}_\nu \left( F k^\rho k^\mu \right) - \bar{\nabla}^\rho \left( F k^\mu k_\nu \right) \right]
\end{splitequation}
with no higher-order terms, provided that $k^\mu$ is geodesic: $k^\rho \bar{\nabla}_\rho k^\mu = 0$. Therefore, if the background metric satisfies the Einstein equations in vacuum $\bar{R}_{\mu\nu} = 0$, one obtains a new solution $R_{\mu\nu} = 0$ by solving the differential equation for $k_\mu$. The Kerr--Schild ansatz can be generalized to metrics which solve the Einstein equations with a cosmological constant, the coupled Einstein--Maxwell system or the Einstein equations with a perfect fluid source~\cite{debneyetal1969,stephanietal2003}, and examples include the charged, rotating black hole solution (and its uncharged and non-rotating limits) and pp-wave spacetimes that describe gravitational radiation.

Since there has been a renewed interest in alternative descriptions of gravity, namely in teleparallel form where gravity is described by torsion instead of curvature~\cite{hehletal1994,obukhovetal1996,obukhovpereira2002,krssaketal2018}, it remains to see if the Kerr--Schild ansatz can be adopted to these situations. In particular, while in Einstein gravity the relevant connection is the Levi-Civita one, which is uniquely determined by the metric such that the above ansatz is all that is needed to determine the relation between the background and full curvature tensors, the teleparallel equivalent employs the tetrad as basic variable and an auxiliary curvature-free (Weitzenböck) connection. We show in section~\ref{sec_kerrschild} that one can generalise the Kerr--Schild ansatz to the tetrad, and derive the analogue of relation~\eqref{kerrschild_ricci} for the torsion. In section~\ref{sec_bh}, we give the solution for a rotating black hole in this framework, and conclude in section~\ref{sec_conclusion}.

\section{The Kerr--Schild ansatz in teleparallel gravity}
\label{sec_kerrschild}

In teleparallel gravity, the basic field variable is the tetrad (or vielbein or frame field) $e_\mu{}^a$, from which the metric is constructed via $g_{\mu\nu} = \eta_{ab} e_\mu{}^a e_\nu{}^b$ with the frame metric $\eta_{ab}$. The inverse of the tetrad is written $e^\mu{}_a$, and fulfills the consistency conditions $e^\mu{}_a = g^{\mu\nu} \eta^{ab} e_\nu{}^b$ and $g^{\mu\nu} = \eta^{ab} e^\mu{}_a e^\nu{}_b$ with the inverse of the frame metric $\eta^{ab}$ and the inverse of the metric $g^{\mu\nu}$. The second field is a spin connection $\omega_{\mu ab} = \omega_{\mu [ab]}$ with vanishing curvature $R_{\mu\nu ab} = \partial_\mu \omega_{\nu ab} - \partial_\nu \omega_{\mu ab} + \omega_{\mu a}{}^c \omega_{\nu cb} - \omega_{\nu a}{}^c \omega_{\mu cb} = 0$, which therefore has the general form~\cite{golovnevetal2017,krssaketal2018}
\begin{equation}
\label{omega_form}
\omega_{\mu ab} = \Lambda^{-1}_{ac} \partial_\mu \Lambda^c{}_b = \Lambda_{ca} \, \partial_\mu \Lambda^c{}_b \eqend{,} \quad \Lambda^a{}_b = \exp( \lambda )^a{}_b
\end{equation}
for an antisymmetric matrix $\lambda_{ab} = \lambda_{[ab]}$ (i.e, it is the Lorentz transformation with parameter $\lambda^a{}_b$ of the zero connection). From this, one obtains the affine connection
\begin{equation}
\label{affine_connection}
\Gamma^\rho_{\mu\nu} \equiv e^\rho{}_a \left( \partial_\mu e_\nu{}^a + \omega_\mu{}^a{}_b \, e_\nu{}^b \right) \eqend{,}
\end{equation}
which for the case of vanishing spin connection $\omega_{\mu ab} = 0$ is known as the Weitzenb{\"o}ck connection. Lastly, one defines torsion $T_{\mu\nu}{}^a$ by
\begin{equation}
T_{\mu\nu}{}^a \equiv 2 \partial_{[\mu} e_{\nu]}{}^a + 2 \omega_{[\mu}{}^{ab} e_{\nu]b} = 2 e_\rho{}^a \Gamma^\rho_{[\mu\nu]} \eqend{.}
\end{equation}
The contortion is the linear combination
\begin{equation}
K_{\mu\rho\sigma} \equiv \frac{1}{2} T_{\rho\sigma\mu} - T_{\mu[\rho\sigma]} \eqend{,}
\end{equation}
which is useful because the connection can be decomposed as
\begin{equation}
\Gamma^\rho_{\mu\nu} = \overring\Gamma^\rho_{\mu\nu} - K_{\mu\nu}{}^\rho \eqend{,} \quad \omega_\mu{}^{ab} = \overring\omega_\mu{}^{ab} + K_{\mu\rho\sigma} e^{\rho a} e^{\sigma b} \eqend{,}
\end{equation}
where $\overring{\Gamma}^\rho_{\mu\nu}$ is the Levi-Civita connection, the unique connection that is both metric-compatible and torsion-free.\footnote{We use the conventions of~\cite{freedmanvanproeyen}. To convert into the conventions used in many of the recent literature on teleparallel gravity, in particular the review~\cite{krssaketal2018} whose teleparallel quantities are denoted with a black bullet, use $\Gamma^\rho_{\mu\nu} = \overbullet\Gamma^\rho_{\nu\mu}$, $\omega_{\mu ab} = \overbullet\omega_{ab\mu}$, $T_{\mu\nu}{}^\rho = \overbullet T^\rho{}_{\mu\nu}$, $K_{\mu\nu}{}^\rho = - \overbullet K{}^\rho{}_{\nu\mu}$ and $S_\rho{}^{\mu\nu} = \overbullet S{}_\rho{}^{\mu\nu}$.}

\subsection{The Kerr--Schild ansatz}

Consider now a background geometry, whose quantities are denoted with an overbar, and a null vector $k_\mu$ in that geometry, fulfilling $\bar{g}^{\mu\nu} k_\mu k_\nu = 0$. Defining $k^\mu \equiv \bar{g}^{\mu\nu} k_\nu$ and $k^a \equiv \bar{e}_\mu{}^a k^\mu$, it is straightforward to check that the ansatz for the full tetrad
\begin{equation}
e_\mu{}^a \equiv \bar{e}_\mu{}^a + F k_\mu k^a
\end{equation}
results in the usual Kerr--Schild form $g_{\mu\nu} = \bar{g}_{\mu\nu} + 2 F k_\mu k_\nu$ for the metric (and also its inverse), that the inverse full tetrad is given by $e^\mu{}_a = \bar{e}^\mu{}_a - F k^\mu k_a$, and that the consistency conditions for the tetrad and the metric hold. Furthermore, the indices of the vector $k_\mu$ can be moved with either the background or full metric and converted into frame indices with either the background or full tetrad, and $k_\mu$ is null also with respect to the full geometry. As in the metric case, the fact that $k_\mu$ is null leads to the equality of the determinants: $e = \det e_\mu{}^a = \det \bar{e}_\mu{}^a = \bar{e}$.

Since the spin connection in teleparallel gravity only represents inertial effects that are present in non-inertial frames~\cite{krssaketal2018}, we restrict to the case where it is equal to the background one: $\omega_{\mu ab} = \bar{\omega}_{\mu ab}$. The physical interpretation is that the Kerr--Schild term results from the addition of gravity, and that the character of the given frame (inertial or not) does not change in the process. For the change of the affine connection, the torsion and the contortion one computes that
\begin{equations}
\Gamma^\rho_{\mu\nu} &= \bar{\Gamma}^\rho_{\mu\nu} + \bar{\nabla}_\mu \left( F k_\nu k^\rho \right) \eqend{,} \\
T_{\mu\nu}{}^a &= \bar{T}_{\mu\nu}{}^a + 2 \bar{\nabla}_{[\mu} \left( F k_{\nu]} k^a \right) + F \bar{T}_{\mu\nu}{}^b k^a k_b \eqend{,} \label{torsion_kerrschild} \\
\begin{split}
K_{\mu\rho\sigma} &= \bar{K}_{\mu\rho\sigma} + 2 \bar{\nabla}_{[\rho} \left( F k_{\sigma]} k_\mu \right) + F \bar{T}_{\rho\sigma}{}^a k_\mu k_a \\
&\quad- 2 F \bar{T}_{\mu[\rho}{}^a k_{\sigma]} k_a \eqend{,}
\end{split}
\end{equations}
where $\bar{\nabla}$ is the covariant derivative of the background geometry involving the spin connection $\omega_{\mu ab}$ for frame indices and the affine connection $\bar{\Gamma}^\rho_{\mu\nu}$ for spacetime indices. To derive these results, one needs to be careful in that the spacetime indices of quantities defined in the full geometry need to be raised or lowered with the full metric $g_{\mu\nu}$, while the ones of background quantities are raised or lowered with the background metric $\bar{g}_{\mu\nu}$, and similarly for frame indices.

\subsection{Equations of motion for teleparallel gravity}

The Lagrangian of the teleparallel equivalent of Einstein gravity is given by~\cite{golovnevetal2017,krssaketal2018}
\begin{splitequation}
\label{lagrangian}
\mathcal{L} &= \frac{\abs{e}}{16 \pi G} \left[ \frac{1}{4} T_{\mu\nu\rho} T^{\mu\nu\rho} + \frac{1}{2} T_{\mu\nu\rho} T^{\mu\rho\nu} - T_\mu T^\mu \right] \\
&= \frac{\abs{e}}{32 \pi G} \mathcal{T}
\end{splitequation}
with the trace of the torsion $T_\mu \equiv T^\rho{}_{\mu\rho}$, the torsion scalar $\mathcal{T}$ and the superpotential $S_\rho{}^{\mu\nu}$:
\begin{equation}
\label{torsionscalar_and_superpotential_def}
\mathcal{T} \equiv S_{\rho\mu\nu} T^{\mu\nu\rho} \eqend{,} \quad S_\rho{}^{\mu\nu} \equiv K_\rho{}^{\mu\nu} - 2 \delta_\rho^{[\mu} T^{\nu]} \eqend{.}
\end{equation}
The variation of the action with respect to the spin connection~\eqref{omega_form}, i.e., with respect to the antisymmetric matrix $\lambda_{ab}$, vanishes identically~\cite{golovnevetal2017,krssaketal2018}. On the other hand, the variation with respect to the tetrad results in
\begin{splitequation}
\label{teleparallel_eom}
&\nabla_\nu S_\rho{}^{\mu\nu} - T_\nu S_\rho{}^{\mu\nu} + T_{\nu\rho}{}^\sigma S_\sigma{}^{\mu\nu} - \frac{1}{2} T_{\nu\sigma}{}^\mu S_\rho{}^{\sigma\nu} \\
&\quad+ \frac{1}{4} \delta_\rho^\mu T_{\alpha\beta}{}^\nu S_\nu{}^{\alpha\beta} = 0 \eqend{.}
\end{splitequation}
Expressing the superpotential in terms of the contortion, this gives
\begin{splitequation}
&\nabla_\nu K_\rho{}^{\mu\nu} + \nabla_\rho T^\mu + T_{\nu\rho}{}^\sigma K_\sigma{}^{\mu\nu} - \frac{1}{2} T_{\nu\sigma}{}^\mu K_\rho{}^{\sigma\nu} - T_\nu K_\rho{}^{\mu\nu} \\
&\quad- \delta_\rho^\mu \left( \nabla_\nu T^\nu - \frac{1}{4} T_{\alpha\beta}{}^\nu K_\nu{}^{\alpha\beta} - \frac{1}{2} T_\nu T^\nu \right) = 0 \eqend{,}
\end{splitequation}
and upon taking the trace in $\rho\mu$, the terms in parentheses in the second line are seen to vanish separately for all spacetime dimensions $n > 2$, to which we restrict in the following. We can thus take only the terms in the first line, and expressing the contortion using the torsion it follows that
\begin{splitequation}
&2 \nabla_\nu T^\nu{}_{(\mu\rho)} + \nabla_\nu T_{\rho\mu}{}^\nu - 2 \nabla_\rho T_\mu + 2 T_{\mu\nu\sigma} T_\rho{}^{(\nu\sigma)} \\
&\quad- \frac{1}{2} T_{\nu\sigma\mu} T^{\nu\sigma}{}_\rho - T^\nu T_{\rho\mu\nu} - 2 T^\nu T_{\nu(\mu\rho)} = 0 \eqend{.}
\end{splitequation}
The part of this equation that is antisymmetric in $\rho\mu$ reads
\begin{equation}
\label{bianchi_torsion}
\nabla_\nu T_{\rho\mu}{}^\nu - 2 \nabla_{[\rho} T_{\mu]} - T^\nu T_{\rho\mu\nu} = 0 \eqend{,}
\end{equation}
which is exactly the contracted Bianchi identity for torsion (since the curvature vanishes)~\cite{freedmanvanproeyen}. We can thus consider only the symmetric part
\begin{splitequation}
\label{eom_full_torsion}
&\nabla_\rho T^\rho{}_{(\mu\nu)} - \nabla_{(\mu} T_{\nu)} + T_{\mu\rho\sigma} T_\nu{}^{(\rho\sigma)} \\
&\quad- \frac{1}{4} T_{\rho\sigma\mu} T^{\rho\sigma}{}_\nu - T^\rho T_{\rho(\mu\nu)} = 0 \eqend{.}
\end{splitequation}

Assume now that the Kerr--Schild null vector $k_\mu$ fulfills the geodesic equation of the background, which for a connection which torsion reads
\begin{equation}
\label{geodesic_eom_torsion}
k^\rho \bar{\nabla}_\rho k^\alpha - \bar{T}^{\alpha\rho\sigma} k_\rho k_\sigma = 0 \eqend{.}
\end{equation}
We assume furthermore that the background torsion $\bar{T}_{\mu\nu\rho}$ fulfills the equation of motion~\eqref{eom_full_torsion}. Raising the $\mu$ index of equation~\eqref{eom_full_torsion} and inserting the Kerr--Schild form, this results in the equation
\begin{splitequation}
&\left( \bar{\nabla}^\rho - \bar{T}^\rho \right) \left[ \bar{\nabla}_\rho \left( k^\mu k_\nu F \right) - \bar{\nabla}^\mu \left( k_\nu k_\rho F \right) - \bar{\nabla}_\nu \left( k^\mu k_\rho F \right) \right] \\
&+ \bar{\nabla}_\rho \left[ \left( \bar{T}^\mu{}_{\sigma\nu} k^\rho + \bar{T}_{\nu\sigma}{}^\mu k^\rho - \bar{T}^{\mu\rho\sigma} k_\nu - \bar{T}_{\nu\rho\sigma} k^\mu \right) k^\sigma F \right] \\
&+ \bar{T}^{\mu\rho\sigma} \bar{\nabla}_\nu \left( k_\rho k_\sigma F \right) - \bar{T}^{\rho\sigma\mu} \bar{\nabla}_\rho \left( k_\nu k_\sigma F \right) \\
&+ \bar{T}_{\nu\rho\sigma} \bar{\nabla}^\mu \left( k^\rho k^\sigma F \right) - \bar{T}_{\rho\sigma\nu} \bar{\nabla}^\rho \left( k^\mu k^\sigma F \right) \\
&- \bar{g}^{\mu\tau} \Big[ \delta^\rho_{(\tau} \bar{T}^{\alpha\beta}{}_{\nu)} \bar{T}_{\alpha\beta\sigma} - 2 \bar{T}^\alpha \delta^\rho_{(\tau} \bar{T}_{\nu)\alpha\sigma} \\
&\ - \bar{T}^{\alpha\rho}{}_\tau \bar{T}_{\alpha\sigma\nu} - 2 \bar{T}^\rho \bar{T}_{\sigma(\tau\nu)} + 4 \bar{T}_{(\tau}{}^{(\rho\alpha)} \bar{T}_{\nu)[\sigma\alpha]} \Big] F k_\rho k^\sigma = 0 \eqend{,} \raisetag{5em}
\end{splitequation}
which as for Einstein gravity is linear in $F k_\mu k_\nu$. In particular, for a flat background with $\bar{T}_{\mu\nu\rho} = 0$ we have the simple second-order equation
\begin{equation}
\label{eom_k_torsion_flat}
\bar{\nabla}^\rho \left[ \bar{\nabla}_\rho \left( k^\mu k_\nu F \right) - \bar{\nabla}^\mu \left( k_\nu k_\rho F \right) - \bar{\nabla}_\nu \left( k^\mu k_\rho F \right) \right] = 0 \eqend{,}
\end{equation}
together with the geodesic equation
\begin{equation}
\label{eom_k_geodesic_flat}
k^\rho \bar{\nabla}_\rho k^\alpha = 0 \eqend{.}
\end{equation}
Note that in this case the equation is the same as in Einstein gravity with a flat background $\bar{R}_{\mu\nu} = 0$~\eqref{kerrschild_ricci}.

\subsection{$f(\mathcal{T})$ teleparallel gravities}

For the case of a flat background with vanishing torsion, the geodesic equation~\eqref{eom_k_geodesic_flat} furthermore implies that the torsion scalar~\eqref{torsionscalar_and_superpotential_def} of the full geometry vanishes:
\begin{equation}
\label{torsionscalar_flat_in_k}
\mathcal{T} = K_{\rho\mu\nu} T^{\mu\nu\rho} - 2 T_\nu T^\nu = - 4 F^2 \left( k^\mu \bar{\nabla}_\mu k^\rho \right) \left( k^\nu \bar{\nabla}_\nu k_\rho \right) = 0 \eqend{.}
\end{equation}
This observation is important for so-called $f(\mathcal{T})$ gravities~\cite{krssaksaridakis2015}, where with a given function $f$ the Lagrangian~\eqref{lagrangian} is generalised to
\begin{equation}
\mathcal{L} = \frac{\abs{e}}{32 \pi G} f(\mathcal{T}) \eqend{.}
\end{equation}
A general variation of this Lagrangian reads
\begin{splitequation}
\delta \mathcal{L} &= \frac{\abs{e}}{32 \pi G} f(\mathcal{T}) \, e^\mu{}_a \delta e_\mu{}^a + \frac{\abs{e}}{32 \pi G} f'(\mathcal{T}) \Big[ - 4 S_a{}^{\mu\nu} \partial_\nu \delta e_\mu{}^a \\
&+ 4 S^{b\mu\nu} \omega_{\nu ab} \delta e_\mu{}^a + 4 S^{a\mu b} \delta \omega_{\mu ab} \Big] + \frac{\abs{e}}{32 \pi G} f'(\mathcal{T}) \delta e_\mu{}^a \\
&\times \left[ - 4 T^{\mu\rho\sigma} T_{a(\rho\sigma)} + 4 T_a T^\mu + 4 T^\rho T_{a\rho}{}^\mu + 2 T^{\rho\nu\mu} T_{a\rho\nu} \right] \eqend{,}
\end{splitequation}
and following~\cite{forgerroemer2004} we decompose the variation of the vierbein into symmetric and antisymmetric parts:
\begin{equation}
\delta e_\mu{}^a = \frac{1}{2} g^{\nu\rho} e_\nu{}^a \delta g_{\mu\rho} - e_\mu{}^b \delta \lambda^a{}_b
\end{equation}
with
\begin{equation}
\delta \lambda^a{}_b \equiv \frac{1}{2} \eta_{bc} g^{\mu\nu} \left( e_\mu{}^a \delta e_\nu{}^c - e_\mu{}^c \delta e_\nu{}^a \right) \eqend{.}
\end{equation}
Integrating by parts, the metric variation then gives directly the symmetric generalisation of the equation of motion~\eqref{teleparallel_eom}
\begin{splitequation}
\label{ft_eom_sym}
&0 = \nabla_\rho \left[ f'(\mathcal{T}) S^{(\mu\nu)\rho} \right] + \frac{1}{4} f(\mathcal{T}) g^{\mu\nu} \\
&\quad- f'(\mathcal{T}) \left[ T_\rho S^{(\mu\nu)\rho} - T^{\rho(\mu|\sigma|} S_\sigma{}^{\nu)\rho} + \frac{1}{2} T_{\rho\sigma}{}^{(\mu} S^{\nu)\sigma\rho} \right] \eqend{.} \raisetag{4.2em}
\end{splitequation}

On the other hand, using the result~\eqref{omega_form} the variation of the spin connection reads
\begin{equation}
\delta \omega_{\mu ab} = \partial_\mu \delta \lambda_{ab} - \omega_{\mu b}{}^c \delta \lambda_{ca} + \omega_{\mu a}{}^c \delta \lambda_{cb} = \nabla_\mu \delta \lambda_{ab} \eqend{,}
\end{equation}
and the $\lambda$ variation of the Lagrangian results in
\begin{splitequation}
&0 = S^{[ab]\mu} f''(\mathcal{T}) \nabla_\mu \mathcal{T} + f'(\mathcal{T}) \bigg[ \nabla_\mu S^{[ab]\mu} \\
&\quad- T_\mu \left( S^{[ab]\mu} + \frac{1}{2} T^{\mu[ab]} \right) - \frac{1}{4} T_{\mu\nu}{}^{[a} \left( S^{b]\mu\nu} + T^{b]\mu\nu} \right) \bigg] \eqend{.} \raisetag{4.6em}
\end{splitequation}
Expressing the superpotential in terms of the torsion, the terms in brackets read
\begin{equation}
- \frac{1}{2} e_\mu{}^a e_\nu{}^b \left[ \nabla_\rho T^{\mu\nu\rho} - 2 \nabla^{[\mu} T^{\nu]} - T_\rho T^{\mu\nu\rho} \right] \eqend{,}
\end{equation}
which is again exactly the contracted Bianchi identity for torsion~\eqref{bianchi_torsion} and vanishes. We see that we obtain a second equation of motion only when the second derivative of $f$ is non-vanishing, which is well known~\cite{krssaksaridakis2015,hohmannetal2019,krssaketal2018}. Expressing the superpotential in terms of the torsion also in the first term, we thus have the equation
\begin{equation}
\label{ft_eom_antisym}
\left( T^{ab\mu} - 2 e^{\mu[a} T^{b]} \right) f''(\mathcal{T}) \nabla_\mu \mathcal{T} = 0 \eqend{.}
\end{equation}

For the Kerr--Schild ansatz (with a flat background), the second equation of motion~\eqref{ft_eom_antisym} is automatically fulfilled since the torsion scalar vanishes according to~\eqref{torsionscalar_flat_in_k}. Moreover, we have to impose $f(0) = 0$, i.e., no cosmological constant, in order to have the flat background $\bar{g}_{\mu\nu} = \eta_{\mu\nu}$ as solution of the first equation of motion~\eqref{ft_eom_sym}. However, then this equation reduces to the one for the teleparallel equivalent of Einstein gravity~\eqref{eom_full_torsion} since the torsion scalar and thus $f'(\mathcal{T})$ is constant. It follows that any solution to the equations~\eqref{eom_k_torsion_flat} and~\eqref{eom_k_geodesic_flat} in the teleparallel equivalent of Einstein gravity with a flat background is also a solution in $f(\mathcal{T})$ gravity with $f(0) = 0$.

\section{The rotating black hole solution}
\label{sec_bh}

In Einstein gravity, the rotating black hole metric was found by Kerr~\cite{kerr1963}, in fact in Kerr--Schild form. The background metric is the flat Minkowski metric $\bar{g}_{\mu\nu} = \eta_{\mu\nu}$ in Cartesian coordinates $x$, $y$, $z$, $t$, the function $F$ is given by
\begin{equation}
\label{bh_F}
F = \frac{M r^3}{r^4 + a^2 z^2}
\end{equation}
with the mass parameter $M = G m$, and the null vector $k^\mu$ is given as the one-form
\begin{splitequation}
\label{bh_kmu}
k_\mu \total x^\mu &= \frac{r}{r^2 + a^2} \left( x \total x + y \total y \right) + \frac{a}{r^2 + a^2} \left( x \total y - y \total x \right) \\
&\quad+ \frac{z}{r} \total z + \total t \eqend{,} \raisetag{2em}
\end{splitequation}
where $r > 0$ is defined as that solution of
\begin{equation}
\label{radius_def}
r^4 - \left( x^2 + y^2 + z^2 - a^2 \right) r^2 - a^2 z^2 = 0
\end{equation}
which asymptotically behaves like the Cartesian radius: $r \sim \sqrt{x^2 + y^2 + z^2}$.

For a flat background the equation determining the null vector $k^\mu$ is the same in Einstein gravity~\eqref{kerrschild_ricci} as in teleparallel gravity~\eqref{eom_k_torsion_flat}, and also the condition that $k^\mu$ be geodesic on the flat background results in the same equation~\eqref{eom_k_geodesic_flat}. It follows immediately that the rotating black hole solution in teleparallel gravity is given in Kerr--Schild form with the same function $F$~\eqref{bh_F} and the same null vector $k^\mu$~\eqref{bh_kmu}, in Cartesian coordinates with the trivial background tetrad $\bar{e}_\mu{}^a = \delta_\mu^a$ and vanishing spin connection $\omega_{\mu ab} = 0$. It thus only remains to compute the torsion explicitly. However, the computation is more meaningful in a frame adapted to the symmetries of the solution. We thus choose as coordinates the time $t$, the radius $r$ defined by~\eqref{radius_def} and two angles $\theta$ and $\phi$ such that
\begin{splitequation}
\label{coord_change}
x &= \sqrt{r^2 + a^2} \sin \theta \cos \phi \eqend{,} \\
y &= \sqrt{r^2 + a^2} \sin \theta \sin \phi \eqend{,} \quad z = r \cos \theta \eqend{.}
\end{splitequation}
These coordinates are known as oblate spheroidal coordinates, where the deviation from spherical form is measured by the parameter $a$, and the background metric reads
\begin{splitequation}
\bar{g}_{\mu\nu} \total x^\mu \total x^\nu &= - \total t^2 + \frac{\Sigma}{r^2 + a^2} \total r^2 \\
&\quad+ \Sigma \total \theta^2 + (r^2 + a^2) \sin^2 \theta \total \phi^2 \eqend{,}
\end{splitequation}
where we set $\Sigma \equiv r^2 + a^2 \cos^2 \theta$. The null vector $k^\mu$ and function $F$ are given by
\begin{equations}[nullvector_func_spherical]
k_\mu \total x^\mu &= \total t + \frac{\Sigma}{r^2 + a^2} \total r + a \sin^2 \theta \total \phi \eqend{,} \\
F &= \frac{M r}{\Sigma} \eqend{,}
\end{equations}
and the full metric reads
\begin{splitequation}
\label{kerrmetric_ks}
&g_{\mu\nu} \total x^\mu \total x^\nu = - \left( 1 - \frac{2 M r}{\Sigma} \right) \total t^2 + \frac{4 M r}{r^2 + a^2} \total t \total r \\
&\quad+ \frac{4 M a r \sin^2 \theta}{\Sigma} \total t \total \phi + \frac{\Sigma}{r^2 + a^2} \left( 1 + \frac{2 M r}{r^2 + a^2} \right) \total r^2 \\
&\quad+ \frac{4 M a r \sin^2 \theta}{r^2 + a^2} \total r \total \phi + \Sigma \total \theta^2 \\
&\quad+ \left( r^2 + a^2 + \frac{2 M a^2 r \sin^2 \theta}{\Sigma} \right) \sin^2 \theta \total \phi^2 \eqend{.} \raisetag{2em}
\end{splitequation}

\subsection{The background geometry}

Since the background geometry should stay flat (and thus have vanishing torsion), the change of coordinates~\eqref{coord_change} also induces a change in the spin connection, which shows explicitly that we have passed from the Cartesian inertial frame to a non-inertial one. Let us choose the background tetrad one-form $\bar{e}^a \equiv \bar{e}_\mu{}^a \total x^\mu$ as
\begin{splitequation}
\label{background_tetrad}
\bar{e}^0 &= \total t \eqend{,} \quad \bar{e}^1 = \sqrt{ \frac{\Sigma}{r^2 + a^2} } \total r \eqend{,} \quad \bar{e}^2 = \sqrt{ \Sigma } \total \theta \eqend{,} \\
\bar{e}^3 &= \sqrt{ r^2 + a^2 } \, \sin \theta \total \phi \eqend{,}
\end{splitequation}
such that the background frame metric has the usual constant form $\bar{\eta}_{ab} = \operatorname{diag}(-1,1,1,1)_{ab}$. Imposing vanishing torsion $0 = \bar{T}_{\mu\nu}{}^\rho = 2 \bar{\Gamma}^\rho_{[\mu\nu]}$, we obtain from equation~\eqref{affine_connection} that
\begin{equation}
\label{background_spin_conn}
\bar{\omega}_{[\mu}{}^{ab} \bar{e}_{\nu]b} = - \partial_{[\mu} \bar{e}_{\nu]}{}^a \eqend{,}
\end{equation}
from which we can determine the background spin connection. Since both sides are antisymmetric in $\mu\nu$, we multiply by $\total x^\mu \wedge \total x^\nu$ to obtain a two-form, whence equation~\eqref{background_spin_conn} reduces to the first Cartan structure equation (for vanishing torsion)
\begin{equation}
\label{background_spin_conn_eq}
\bar{\omega}^{ab} \wedge \bar{e}_b = - \total \bar{e}^a \eqend{,}
\end{equation}
where $\total \bar{e}^a \equiv \partial_\mu \bar{e}_\nu{}^a \total x^\mu \wedge \total x^\nu$, and $\bar{\omega}^{ab} \equiv \bar{\omega}_{\mu}{}^{ab} \total x^\mu$ is the background spin connection one-form. For the right-hand side, we compute that
\begin{splitequation}
\total \bar{e}^0 &= 0 \eqend{,} \quad \total \bar{e}^1 = \frac{a^2 \cos \theta \sin \theta}{\sqrt{ (r^2 + a^2) \Sigma }} \total r \wedge \total \theta \eqend{,} \\
\total \bar{e}^2 &= \frac{r}{\sqrt{ \Sigma }} \total r \wedge \total \theta \eqend{,} \\
\total \bar{e}^3 &= \frac{r \sin \theta}{\sqrt{ r^2 + a^2 }} \total r \wedge \total \phi + \sqrt{ r^2 + a^2 } \, \cos \theta \total \theta \wedge \total \phi \eqend{.}
\end{splitequation}

Imposing the antisymmetry $\bar{\omega}^{ab} = \bar{\omega}^{[ab]}$, equation~\eqref{background_spin_conn_eq} then results in
\begin{equations}
\begin{split}
0 &= \sqrt{ \frac{\Sigma}{r^2 + a^2} } \, \bar{\omega}^{01} \wedge \total r + \sqrt{ \Sigma } \, \bar{\omega}^{02} \wedge \total \theta \\
&\quad+ \sqrt{ r^2 + a^2 } \, \sin \theta \, \bar{\omega}^{03} \wedge \total \phi \eqend{,}
\end{split} \\
\begin{split}
0 &= \bar{\omega}^{01} \wedge \total t + \sqrt{ r^2 + a^2 } \, \sin \theta \, \bar{\omega}^{13} \wedge \total \phi \\
&\quad+ \sqrt{ \Sigma } \, \bar{\omega}^{12} \wedge \total \theta + \frac{a^2 \cos \theta \sin \theta}{\sqrt{ (r^2 + a^2) \Sigma }} \total r \wedge \total \theta \eqend{,}
\end{split} \\
\begin{split}
0 &= \bar{\omega}^{02} \wedge \total t - \sqrt{ \frac{\Sigma}{r^2 + a^2} } \, \bar{\omega}^{12} \wedge \total r \\
&\quad+ \sqrt{ r^2 + a^2 } \, \sin \theta \, \bar{\omega}^{23} \wedge \total \phi + \frac{r}{\sqrt{ \Sigma }} \total r \wedge \total \theta \eqend{,}
\end{split} \\
\begin{split}
0 &= \bar{\omega}^{03} \wedge \total t - \sqrt{ \frac{\Sigma}{r^2 + a^2} } \, \bar{\omega}^{13} \wedge \total r - \sqrt{ \Sigma } \, \bar{\omega}^{23} \wedge \total \theta \\
&\quad+ \frac{r \sin \theta}{\sqrt{ r^2 + a^2 }} \total r \wedge \total \phi + \sqrt{ r^2 + a^2 } \, \cos \theta \total \theta \wedge \total \phi \eqend{,}
\end{split}
\end{equations}
which after some algebra gives the unique result for the background spin connection
\begin{splitequation}
\label{background_spin_conn_form}
\bar{\omega}^{0b} &= 0 \eqend{,} \quad \bar{\omega}^{12} = - \frac{a^2 \cos \theta \sin \theta}{\Sigma \sqrt{ r^2 + a^2 }} \total r - \frac{r \sqrt{r^2 + a^2}}{\Sigma} \total \theta \eqend{,} \\
\bar{\omega}^{13} &= - \frac{r \sin \theta}{\sqrt{ \Sigma }} \total \phi \eqend{,} \quad \bar{\omega}^{23} = - \sqrt{ \frac{r^2 + a^2}{\Sigma} } \, \cos \theta \total \phi \eqend{.}
\end{splitequation}
Since the background torsion vanishes, this is actually equal to the Levi-Civita connection, and in fact the connection naturally associated to the given background te\-trad~\cite{krssakpereira2015}.

It is important to note that the background geometry given by the tetrad~\eqref{background_tetrad} and the spin connection~\eqref{background_spin_conn_form} is flat (i.e., has vanishing torsion $\bar{T}_{\mu\nu}{}^a = 0$) for all values of the parameter $a$, which at this point does not yet have a physical interpretation. Moreover, even as $a \to 0$ where one recovers spherical coordinates we have a non-inertial frame where the spin connection does not vanish. (In this limit, it corresponds to the reference tetrad of~\cite[eq.~(5.8)]{krssakpereira2015} or the orthogonal tetrad of~\cite[eq.~(78)]{hohmannetal2019} with $C_1 = C_4 = 1$, $C_s = r$ and $C_2 = C_3 = C_\alpha = 0$.)

\subsection{The Kerr--Schild geometry}

It is now a long but straightforward computation to check that the null vector $k^\mu$ fulfills the geodesic equation~\eqref{eom_k_geodesic_flat} and the equation of motion~\eqref{eom_k_torsion_flat}, which is somewhat simplified by the fact that in the adopted frame everything only depends on $r$ and $\theta$. The geodesic equation can be rewritten as
\begin{equation}
k^\rho \partial_\rho k^a + k^\rho \, \bar{\omega}_\rho{}^{ab} k_b = 0 \eqend{,}
\end{equation}
where the inverse and frame components of the null vector are given by
\begin{equations}
k^t &= - 1 \eqend{,} \quad k^r = 1 \eqend{,} \quad k^\theta = 0 \eqend{,} \quad k^\phi = \frac{a}{r^2 + a^2} \eqend{,} \\
\begin{split}
\label{k_frame}
k_0 &= 1 = - k^0 \eqend{,} \quad k_1 = \sqrt{ \frac{\Sigma}{r^2 + a^2} } = k^1 \eqend{,} \\
k_2 &= 0 = k^2 \eqend{,} \quad k_3 = \frac{a \sin \theta}{\sqrt{ r^2 + a^2 }} = k^3 \eqend{,}
\end{split}
\end{equations}
and is easy to verify. On the other hand, the second-order equation~\eqref{eom_k_torsion_flat} is more difficult to check, and we have verified it using the xCoba package of the xAct tensor algebra suite for Mathematica~\cite{xact}, which was also used for the computation of the torsion.

For the torsion two-form $T^a \equiv \frac{1}{2} T_{\mu\nu}{}^a \total x^\mu \wedge \total x^\nu$, equation~\eqref{torsion_kerrschild} reduces to
\begin{equation}
T^a = \partial_\mu \left( F k_\nu k^a \right) \total x^\mu \wedge \total x^\nu + \bar{\omega}_\mu{}^{ab} F k_\nu k_b \total x^\mu \wedge \total x^\nu \eqend{,}
\end{equation}
which for the individual components is
\begin{equations}
T^0 &= - \partial_\mu \left( k_\nu F \right) \total x^\mu \wedge \total x^\nu \eqend{,} \\
T^1 &= \partial_\mu \left( k_\nu k_1 F \right) \total x^\mu \wedge \total x^\nu + k_\nu k_3 F \bar{\omega}^{13} \wedge \total x^\nu \eqend{,} \\
T^2 &= - k_\nu k_1 F \bar{\omega}^{12} \wedge \total x^\nu + k_\nu k_3 F \bar{\omega}^{23} \wedge \total x^\nu \eqend{,} \\
T^3 &= \partial_\mu \left( k_\nu k_3 F \right) \total x^\mu \wedge \total x^\nu - k_\nu k_1 F \bar{\omega}^{13} \wedge \total x^\nu \eqend{.}
\end{equations}
We compute
\begin{equations}[torsion_result]
\begin{split}
T^0 &= - \frac{M (2 r^2 - \Sigma)}{\Sigma^2} \left( \total t + a \sin^2 \theta \total \phi \right) \wedge \total r \\
&\quad+ \frac{2 M a r \sin \theta \cos \theta}{\Sigma^2} \left[ a \total t + (r^2 + a^2) \total \phi \right] \wedge \total \theta \eqend{,}
\end{split} \raisetag{6em} \\
\begin{split}
T^1 &= \frac{M (r^4 - a^4 \cos^2 \theta)}{\Sigma^\frac{3}{2} (r^2 + a^2)^\frac{3}{2}} \left( \total t + a \sin^2 \theta \total \phi \right) \wedge \total r \\
&\quad+ \frac{M a r \sin \theta}{( r^2 + a^2 )^\frac{3}{2} \sqrt{ \Sigma }} \total r \wedge \left( r \sin \theta \total \phi + a \cos \theta \total \theta \right) \\
&\quad+ \frac{M r \sin \theta}{\Sigma^\frac{3}{2} \sqrt{r^2 + a^2}} \left[ a \total t + (r^2 + a^2 + \Sigma) \total \phi \right] \\
&\qquad\qquad\wedge \left( r \sin \theta \total \phi - a \cos \theta \total \theta \right) \eqend{,}
\end{split} \raisetag{4em} \\
\begin{split}
T^2 &= - \frac{M a^2 r \cos \theta \sin \theta}{(r^2 + a^2) \Sigma^\frac{3}{2}} \left( \total t + a \sin^2 \theta \total \phi \right) \wedge \total r \\
&\quad- \frac{M r}{(r^2 + a^2) \sqrt{ \Sigma }} \total r \wedge \left( r \total \theta - a \sin \theta \cos \theta \total \phi \right) \\
&\quad- \frac{M r}{\Sigma^\frac{3}{2}} \left( \total t + a \sin^2 \theta \total \phi \right) \wedge \left( r \total \theta - a \sin \theta \cos \theta \total \phi \right) \eqend{,}
\end{split} \raisetag{6em} \\
\begin{split}
T^3 &= \frac{M a \sin \theta (2 r^4 + a^2 r^2 - a^4 \cos^2 \theta)}{\Sigma^2 (r^2 + a^2)^\frac{3}{2}} \\
&\qquad\times \left( \total t + a \sin^2 \theta \total \phi \right) \wedge \total r \\
&\quad- \frac{M r a \cos \theta (\Sigma + 2 a^2 \sin^2 \theta)}{\Sigma^2 \sqrt{ r^2 + a^2 }} \left( \total t + a \sin^2 \theta \total \phi \right) \wedge \total \theta \\
&\quad- \frac{M r \sin \theta}{\Sigma \sqrt{ r^2 + a^2 }} \left( r \total t - 2 a^2 \sin \theta \cos \theta \total \theta \right) \wedge \total \phi \\
&\quad- \frac{M r}{(r^2 + a^2)^\frac{3}{2}} \total r \wedge \left( a \cos \theta \total \theta + r \sin \theta \total \phi \right) \eqend{,}
\end{split} \raisetag{3.2em}
\end{equations}
which is unfortunately not very enlightening.

Because the spacetime is algebraically special, the totally antisymmetric part of the torsion has to vanish, as can be seen from the relation~\eqref{torsion_kerrschild}. This also serves as a check of the above computation, and we verify that
\begin{equation}
\label{torsion_antisym}
T_{\mu\nu\rho} \total x^\mu \wedge \total x^\nu \wedge \total x^\rho = 2 \eta_{ab} T^a \wedge \bar{e}^b + 2 F k_a k_\nu T^a \wedge \total x^\nu = 0 \eqend{.}
\end{equation}
Lastly, for the trace of the torsion $T \equiv T_{\mu\nu}{}^a e^\mu{}_a \total x^\nu = \bar{\nabla}_\mu \left( F k_\nu k^\mu \right) \total x^\nu$, we obtain
\begin{equation}
\label{torsion_trace}
T = \frac{M}{r^2 + a^2} \total r + \frac{M}{\Sigma} \left( \total t + a \sin^2 \theta \total \phi \right) = \frac{F}{r} k_\mu \total x^\mu \eqend{.}
\end{equation}
That the trace is proportional to the null vector might seem surprising at first, but actually holds as a consequence of the geodesic equation satisfied by the null vector. Namely, the general result~\eqref{torsion_kerrschild} (with non-vanishing background torsion) gives
\begin{splitequation}
T_\nu &= \bar{T}_\nu + 2 \bar{\nabla}_{[\mu} \left( F k_{\nu]} k^\mu \right) \\
&= \bar{T}_\nu + F \bar{T}_{\nu\rho\sigma} k^\rho k^\sigma + k_\nu \bar{\nabla}_\mu \left( F k^\mu \right) \eqend{,}
\end{splitequation}
where for the second equality we used the geodesic equation with torsion~\eqref{geodesic_eom_torsion}. If the background torsion vanishes, as in our case, we obtain the observed proportionality, and can moreover derive that
\begin{equation}
\label{fkmu_der}
\bar{\nabla}_\mu \left( F k^\mu \right) = k^\mu \partial_\mu F + F \bar{e}^\mu{}_a \partial_\mu k^a + F \bar{e}^\mu{}_a \, \bar{\omega}_\mu{}^a{}_b \, k^b = \frac{F}{r} \eqend{,}
\end{equation}
which can be checked to hold.

\subsection{Non-rotating limit}

In the non-rotating limit $a \to 0$, the background tetrad \eqref{background_tetrad} becomes the standard orthogonal tetrad for spherical symmetry,
\begin{equation}
\label{tetrad_nonrotating}
\bar{e}^0 \to \total t \eqend{,} \quad \bar{e}^1 \to \total r \eqend{,} \quad \bar{e}^2 \to r \total \theta \eqend{,} \quad \bar{e}^3 \to r \sin \theta \total \phi \eqend{,}
\end{equation}
while the spin connection~\eqref{background_spin_conn_form} reduces to
\begin{splitequation}
\label{spinconn_nonrotating}
&\bar{\omega}^{0b} \to 0 \eqend{,} \quad \bar{\omega}^{12} \to - \total \theta \eqend{,} \\
&\bar{\omega}^{13} \to - \sin \theta \total \phi \eqend{,} \quad \bar{\omega}^{23} \to - \cos \theta \total \phi \eqend{,}
\end{splitequation}
both agreeing with the reference tetrad of~\cite[eq.~(5.8)]{krssakpereira2015} and the orthogonal tetrad of~\cite[eq.~(78)]{hohmannetal2019} with $C_1 = C_4 = 1$, $C_s = r$ and $C_2 = C_3 = C_\alpha = 0$ and the corresponding spin connection. The solution for the torsion~\eqref{torsion_result} reduces to
\begin{equations}[torsion_nonrotating]
T^0 &= - \frac{M}{r^2} \total t \wedge \total r \eqend{,} \\
T^1 &= \frac{M}{r^2} \total t \wedge \total r \eqend{,} \\
T^2 &= - \frac{M}{r} \left( \total t + \total r \right) \wedge \total \theta \eqend{,} \\
T^3 &= - \frac{M \sin \theta}{r} \left( \total t + \total r \right) \wedge \total \phi \eqend{,}
\end{equations}
which however does not agree with the Schwarzschild solution of~\cite{emtsovaetal2019}. The reason is the same as for the metric case, namely that the Kerr--Schild form leads to the Schwarzschild solution in coordinates related to the Edd\-ington--Finkelstein ones:
\begin{splitequation}
g_{\mu\nu} \total x^\mu \total x^\nu &= - \left( 1 - \frac{2 M}{r} \right) \total t^2 + \left( 1 + \frac{2 M}{r} \right) \total r^2 \\
&\quad+ \frac{4 M}{r} \total t \total r + r^2 \left( \total \theta^2 + \sin^2 \theta \total \phi^2 \right) \eqend{,}
\end{splitequation}
which in particular contains a mixed term $\total t \total r$.

To recover the solution in Schwarzschild coordinates, we perform the shift
\begin{equation}
t \to t + 2 M \ln \left( r - 2 M \right) \eqend{,}
\end{equation}
which results in
\begin{equation}
\total t \to \total t + \frac{2 M}{r - 2 M} \total r
\end{equation}
and the well-known metric
\begin{splitequation}
g_{\mu\nu} \total x^\mu \total x^\nu &= - \left( 1 - \frac{2 M}{r} \right) \total t^2 + \left( 1 - \frac{2 M}{r} \right)^{-1} \total r^2 \\
&\quad+ r^2 \total \theta^2 + r^2 \sin^2 \theta \total \phi^2 \eqend{.} \raisetag{1.4em}
\end{splitequation}
The shift only affects the $0$ and $1$ components of the full tetrad $e^a = \bar{e}^a + F k^a k_\mu \total x^\mu$, for which we obtain
\begin{equations}
e^0 &= \total t - \frac{M}{r} \left( \total t + \total r \right) \to \left( 1 - \frac{M}{r} \right) \total t + \frac{M}{r - 2 M} \total r \eqend{,} \\
e^1 &= \total r + \frac{M}{r} \left( \total t + \total r \right) \to \frac{M}{r} \total t + \frac{r - M}{r - 2 M} \total r \eqend{,}
\end{equations}
where we used the limit $a \to 0$ of the null vector frame components~\eqref{k_frame}. To be able to compare with the result of~\cite{emtsovaetal2019}, we need to bring this into diagonal form, which is achieved with the local Lorentz transformation $e^a \to \tilde{e}^a = \Lambda^a{}_b e^b = \exp( \lambda )^a{}_b e^b$ with
\begin{equations}[nonrotating_lorentz]
\Lambda^a{}_b &= \begin{pmatrix} \frac{r-M}{\sqrt{ r (r - 2 M) }} & - \frac{M}{\sqrt{ r (r - 2 M) }} & 0 & 0 \\ - \frac{M}{\sqrt{ r (r - 2 M) }} & \frac{r-M}{\sqrt{ r (r - 2 M) }} & 0 & 0 \\ 0 & 0 & 1 & 0 \\ 0 & 0 & 0 & 1 \end{pmatrix} \eqend{,} \\
\lambda_{ab} &= \begin{pmatrix} 0 & - \ln \sqrt{ 1 - \frac{2 M}{r} } & 0 & 0 \\ \ln \sqrt{ 1 - \frac{2 M}{r} } & 0 & 0 & 0 \\ 0 & 0 & 0 & 0 \\ 0 & 0 & 0 & 0 \end{pmatrix} \eqend{.}
\end{equations}
This results in the diagonal tetrad
\begin{splitequation}
\tilde{e}^0 &= \sqrt{ 1 - \frac{2 M}{r} } \total t \eqend{,} \quad \tilde{e}^1 = \left( 1 - \frac{2 M}{r} \right)^{-\frac{1}{2}} \total r \eqend{,} \\
\tilde{e}^2 &= r \total \theta \eqend{,} \quad \tilde{e}^3 = r \sin \theta \total \phi \eqend{,}
\end{splitequation}
which agrees with~\cite{emtsovaetal2019}, while the background spin connection~\eqref{spinconn_nonrotating} is unaffected (and agrees with~\cite{emtsovaetal2019}) since for $M = 0$ the transformation~\eqref{nonrotating_lorentz} is the identity.

Under a general Lorentz transformation, we have
\begin{equation}
\label{torsion_lorentztrafo}
\tilde{T}^a = \Lambda^a{}_b T^b + \left[ - \Lambda^a{}_b \, \omega^{bd} \eta_{dc} + \total \Lambda^a{}_c + \tilde{\omega}^{ab} \Lambda_{bc} \right] \wedge e^c \eqend{.}
\end{equation}
If the connection transforms in the well-known inhomogeneous fashion such that the term in brackets vanishes, the torsion transforms as a true tensor. However, in our case the connection is determined by the background and kept fixed, and consequently the torsion receives an additional piece. We compute
\begin{equations}[torsion_nonrotating_lorentz]
\tilde{T}^0 &= - \frac{M}{r \sqrt{ r (r - 2 M) }} \total t \wedge \total r \eqend{,} \\
\tilde{T}^1 &= 0 \eqend{,} \\
\tilde{T}^2 &= \left[ 1 - \left( 1 - \frac{2 M}{r} \right)^{-\frac{1}{2}} \right] \total r \wedge \total \theta \eqend{,} \\
\tilde{T}^3 &= \sin \theta \left[ 1 - \left( 1 - \frac{2 M}{r} \right)^{-\frac{1}{2}} \right] \total r \wedge \total \phi \eqend{,}
\end{equations}
which then agrees with the Schwarzschild solution of~\cite{emtsovaetal2019} after expressing their spacetime components in terms of tetrad ones.

\subsection{Asymptotic form}

To have a better understanding of the parameters $M$ and $a$ in the solution, we determine the form of all geometric objects as we approach spatial infinity $r \to \infty$. For the background tetrad and the spin connection, this limit is equal to the spherically symmetric one~\eqref{tetrad_nonrotating} and~\eqref{spinconn_nonrotating}, up to terms that are supressed at least by $r^{-2}$ relative to the terms shown. For the components of the torsion~\eqref{torsion_result} and its trace~\eqref{torsion_trace}, we first express the coordinate differentials in terms of the tetrad, and obtain to leading order
\begin{equations}[torsion_asymp]
T^0 &\to \frac{M}{r^2} \left[ - \bar{e}^0 \wedge \bar{e}^1 + \frac{a \sin \theta}{r} \bar{e}^1 \wedge \bar{e}^3 - \frac{2 a \cos \theta}{r} \bar{e}^2 \wedge \bar{e}^3 \right] \eqend{,} \\
T^1 &\to \frac{M}{r^2} \left[ \bar{e}^0 \wedge \bar{e}^1 + \frac{a \sin \theta}{r} \bar{e}^0 \wedge \bar{e}^3 + \frac{2 a \cos \theta}{r} \bar{e}^2 \wedge \bar{e}^3 \right] \eqend{,} \\
\begin{split}
T^2 &\to \frac{M}{r^2} \bigg[ - \bar{e}^0 \wedge \bar{e}^2 - \bar{e}^1 \wedge \bar{e}^2 + \frac{a \cos \theta}{r} \left( \bar{e}^0 + \bar{e}^1 \right) \wedge \bar{e}^3 \\
&\qquad\qquad+ \frac{a \sin \theta}{r} \bar{e}^2 \wedge \bar{e}^3 \bigg] \eqend{,}
\end{split} \raisetag{2em} \\
\begin{split}
T^3 &\to \frac{M}{r^2} \bigg[ \frac{2 a \sin \theta}{r} \bar{e}^0 \wedge \bar{e}^1 - \frac{a \cos \theta}{r} \left( \bar{e}^0 + \bar{e}^1 \right) \wedge \bar{e}^2 \\
&\qquad\qquad- \left( \bar{e}^0 + \bar{e}^1 \right) \wedge \bar{e}^3 \bigg] \eqend{,}
\end{split} \\
T &\to \frac{M}{r^2} \left[ \bar{e}^0 + \bar{e}^1 + \frac{a \sin \theta}{r} \bar{e}^3 \right] \eqend{,}
\end{equations}
where the terms that are not shown are at least of order $r^{-4}$. We see that the leading behaviour at spatial infinity is determined by the parameter $M$, while the next-to-leading one is linear in $a$. In a putative multipole expansion of the torsion similar to the one established for Einstein gravity~\cite{geroch1970,hansen1974,thorne1980,simonbeig1983}, the leading term should correspond to the monopole, which can be interpreted as the total mass of the spacetime, while the next-to-leading one corresponds to the dipole and can be interpreted as the total angular momentum.

While it is possible to compute conserved charges that directly give the total mass and angular momentum~\cite{deandradeetal2000,obukhovrubilar2006,obukhovetal2006,krssakpereira2015,krssaketal2018,emtsovaetal2019}, these again depend quite sensitively on the choice of tetrad and background spin connection. In particular, we can perform a change of coordinates to pass to Boyer--Lindquist ones, which is given by
\begin{equation}
\label{boyerlindquist_shift}
\total t \to \total t + \frac{2 M r}{\Delta} \total r \eqend{,} \quad \total \phi \to \total \phi - \frac{2 M r}{\Delta} \frac{a}{r^2 + a^2} \total r
\end{equation}
with
\begin{equation}
\Delta \equiv r^2 + a^2 - 2 M r \eqend{,}
\end{equation}
and changes the metric~\eqref{kerrmetric_ks} to
\begin{splitequation}
\label{kerrmetric_boyerlindquist}
g_{\mu\nu} \total x^\mu \total x^\nu &\to - \total t^2 + \frac{\Sigma}{\Delta} \total r^2 + (r^2 + a^2) \sin^2 \theta \total \phi^2 \\
&\quad+ \Sigma \total \theta^2 + \frac{2 M r}{\Sigma} \left( \total t + a \sin^2 \theta \total \phi \right)^2 \eqend{.}
\end{splitequation}
This time, all components of the full tetrad are affected, and we compute
\begin{equations}[boyer_lindquist_tetrad]
e^0 &\to \left( 1 - \frac{M r}{\Sigma} \right) \total t + \frac{M r}{\Delta} \total r - \frac{M r a \sin^2 \theta}{\Sigma} \total \phi \eqend{,} \\
\begin{split}
e^1 &\to \frac{M r}{\sqrt{ \Sigma (r^2+a^2) }} \left( \total t + a \sin^2 \theta \total \phi \right) \\
&\quad+ \sqrt{ \frac{\Sigma}{r^2 + a^2} } \frac{\Delta + M r}{\Delta} \total r \eqend{,}
\end{split} \\
e^2 &\to \sqrt{ \Sigma } \total \theta \eqend{,} \\
\begin{split}
e^3 &\to \frac{M r a \sin \theta}{\Sigma \sqrt{r^2 + a^2}} \total t - \frac{M r a \sin \theta}{\Delta \sqrt{r^2 + a^2}} \total r \\
&\quad+ \frac{(r^2 + a^2) \Sigma + M r a^2 \sin^2 \theta}{\Sigma \sqrt{r^2 + a^2}} \sin \theta \total \phi \eqend{.}
\end{split}
\end{equations}

To bring this tetrad into the form used in~\cite{lucasetal2009,krssakpereira2015}, we set $A^2 = \Delta \Sigma + 2 M r ( r^2 + a^2 )$ and $B = \sqrt{ \frac{r^2 + a^2}{\Sigma} }$ and perform a Lorentz transformation with
\begin{equation}
\Lambda^a{}_b = \begin{pmatrix} \frac{\Delta \Sigma + M r ( r^2 + a^2 )}{A \sqrt{\Delta \Sigma}} & - \frac{M r \sqrt{r^2 + a^2}}{A \sqrt{\Delta}} & 0 & \frac{B M a r \sin \theta}{A \sqrt{\Delta}} \\ - \frac{M r}{\sqrt{\Delta \Sigma}} & \frac{\Delta + M r}{\sqrt{\Delta (r^2+a^2)}} & 0 & - \frac{M a r \sin \theta}{\sqrt{\Delta \Sigma (r^2+a^2)}} \\ 0 & 0 & 1 & 0 \\ \frac{M a r \sin \theta}{A \sqrt{\Sigma}} & \frac{M a r \sin \theta}{A \sqrt{r^2+a^2}} & 0 & \frac{\Delta + M r}{A B} + \frac{B M r}{A} \end{pmatrix} \eqend{,}
\end{equation}
which leads to the tetrad $\tilde{e}^a = \Lambda^a{}_b e^b$ with
\begin{splitequation}
\tilde{e}^0 &= \frac{\sqrt{\Delta \Sigma}}{A} \total t \eqend{,} \quad \tilde{e}^1 = \sqrt{\frac{\Sigma}{\Delta}} \total r \eqend{,} \\
\tilde{e}^2 &= \sqrt{\Sigma} \total \theta \eqend{,} \quad \tilde{e}^3 = \frac{2 M a r}{A \sqrt{\Sigma}} \sin \theta \total t + \frac{A}{\sqrt{\Sigma}} \sin \theta \total \phi
\end{splitequation}
and the inverse tetrad
\begin{splitequation}
\tilde{e}^\mu_0 &= \frac{A}{\sqrt{\Delta \Sigma}} \delta^\mu_t - \frac{2 M a r}{A \sqrt{\Delta \Sigma}} \delta^\mu_\phi \eqend{,} \quad \tilde{e}^\mu_1 = \sqrt{\frac{\Delta}{\Sigma}} \delta^\mu_r \eqend{,} \\
\tilde{e}^\mu_2 &= \frac{1}{\sqrt{\Sigma}} \delta^\mu_\theta \eqend{,} \hspace{7.6em} \tilde{e}^\mu_3 = \frac{\sqrt{\Sigma}}{A \sin \theta} \delta^\mu_\phi \eqend{.}
\end{splitequation}
Since the shift~\eqref{boyerlindquist_shift} vanishes for $M = 0$ the background spin connection is unaffected and we keep it again fixed, which leads to an inhomogeneous transformation of the torsion~\eqref{torsion_lorentztrafo}. We content ourselves with the next-to-leading order for large $r$, where
\begin{equation}
\Lambda^a{}_b = \begin{pmatrix} 1 + \frac{M^2}{2 r^2} & - \frac{M (r+M)}{r^2} & 0 & \frac{M a \sin \theta}{r^2} \\ - \frac{M (r+M)}{r^2} & 1 + \frac{M^2}{2 r^2} & 0 & - \frac{M a \sin \theta}{r^2} \\ 0 & 0 & 1 & 0 \\ \frac{M a \sin \theta}{r^2} & \frac{M a \sin \theta}{r^2} & 0 & 1 \end{pmatrix} + \bigo{r^{-3}} \eqend{,}
\end{equation}
and it follows that
\begin{equations}[torsion_lorentz]
\tilde{T}^0 &= - \frac{M}{r^2} \left( 1 + \frac{M}{r} \right) \bar{e}^0 \wedge \bar{e}^1 \eqend{,} \\
\tilde{T}^1 &= \frac{2 M a \sin \theta}{r^3} \bar{e}^0 \wedge \bar{e}^3 \eqend{,} \\
\tilde{T}^2 &= \frac{M}{r^2} \left[ - \bar{e}^1 \wedge \bar{e}^2 - \frac{3 M}{2 r} \bar{e}^1 \wedge \bar{e}^2 + \frac{2 a \cos \theta}{r} \bar{e}^0 \wedge \bar{e}^3 \right] \eqend{,} \\
\begin{split}
\tilde{T}^3 &= \frac{M}{r^2} \bigg[ - \bar{e}^1 \wedge \bar{e}^3 - \frac{3 M}{2 r} \bar{e}^1 \wedge \bar{e}^3 + \frac{4 a \sin \theta}{r} \bar{e}^0 \wedge \bar{e}^1 \\
&\qquad\qquad- \frac{2 a \cos \theta}{r} \bar{e}^0 \wedge \bar{e}^2 \bigg] \eqend{,}
\end{split} \raisetag{2em} \\
\tilde{T} &= \frac{M}{r^2} \left( 1 + \frac{M}{r} \right) \bar{e}^1 \eqend{,}
\end{equations}
where the first neglected terms are of order $r^{-4}$.

With the given symmetry, the conserved charges are obtained from the $tr$ components of the superpotential \eqref{torsionscalar_and_superpotential_def} \cite{obukhovrubilar2006,lucasetal2009,krssakpereira2015,emtsovaetal2019}. For the original tetrad and torsion~\eqref{torsion_asymp}, we compute 
\begin{splitequation}
\abs{e} S_\rho{}^{tr} \total x^\rho &= - 2 M \sin \theta \total t - 2 M \sin \theta \total r \\
&\quad- 2 M a \sin^3 \theta \total \phi + \bigo{r^{-1}} \eqend{,}
\end{splitequation}
while for the transformed tetrad and torsion~\eqref{torsion_lorentz} we obtain
\begin{equation}
\abs{\tilde{e}} \tilde{S}_\rho{}^{tr} \total x^\rho = - 2 M \sin \theta \total t - 3 M a \sin^3 \theta \total \phi + \bigo{r^{-1}} \eqend{,}
\end{equation}
which coincides with~\cite{krssakpereira2015} taking into account their conventions. The corresponding conserved charges are obtained from
\begin{equation}
P_\rho = \frac{1}{8 \pi G} \lim_{r \to \infty} \int \abs{e} S_\rho{}^{tr} \total \theta \total \phi \eqend{,}
\end{equation}
and we compute $P_\rho = (-m,-m,0, \frac{2}{3} m a)$ (using that $M = G m$) and $\tilde{P}_\rho = (-m,0,0,-m a)$, where the latter again agrees with~\cite{krssakpereira2015} taking into account the different conventions, as well as the result in Einstein gravity~\cite{aguirregabiriaetal1995}.

\subsection{Pure tetrad approach}

A different way to fix the ambiguity in the choice of spin connection is to work in a pure tetrad approach with vanishing spin connection, i.e., taking the Weitzenböck connection $\Gamma^\rho_{\mu\nu} = e^\rho{}_a \partial_\mu e_\nu{}^a$ as affine connection. Rotating black holes in the pure tetrad approach in $f(\mathcal{T})$ gravity have been studied in Refs.~\cite{pereiraetal2001,malufetal2002,malufetal2005} and more recently, including electromagnetism, in~\cite{capozzielloetal2012,ahmedetal2016,capozziellonashed2019}, to which we can compare our results.

To obtain a solution with vanishing spin connection, we have to determine a Lorentz transformation $\Lambda^a{}_b$ such that the transformed spin connection vanishes,
\begin{equation}
\label{pure_tetrad_spinconn}
\tilde{\omega}^{ab} \Lambda_{bc} = \Lambda^a{}_b \, \omega^{bd} \eta_{dc} - \total \Lambda^a{}_c = 0 \eqend{,}
\end{equation}
which then also leads to a covariant transformation of the torsion $\tilde{T}^a$~\eqref{torsion_lorentztrafo}. Since the spin connection is fixed to be the background one~\eqref{background_spin_conn_form}, this is not too difficult, and the required Lorentz transformation reads
\begin{equation}
\label{pure_tetrad_lorentz}
\Lambda^a{}_b = \begin{pmatrix} 1 & 0 & 0 & 0 \\ 0 & B \cos \theta & - \frac{r \sin \theta}{\sqrt{\Sigma}} & 0 \\ 0 & \frac{r \sin \theta}{\sqrt{ \Sigma }} \cos \phi & B \cos \theta \cos \phi & - \sin \phi \\ 0 & \frac{r \sin \theta}{\sqrt{ \Sigma }} \sin \phi & B \cos \theta \sin \phi & \cos \phi \end{pmatrix} \eqend{.}
\end{equation}
Using the tetrad in Boyer--Lindquist coordinates~\eqref{boyer_lindquist_tetrad}, the transformed tetrad $\tilde{e}^a = \Lambda^a{}_b e^b$ is given by
\begin{equations}[pure_tetrad_boyerlindquist]
\tilde{e}^0 &= \total t - U \eqend{,} \\
\tilde{e}^1 &= \frac{(r^2 + a^2) \cos \theta}{\Delta} \total r - r \sin \theta \total \theta + \cos \theta \, U \eqend{,} \\
\begin{split}
\tilde{e}^2 &= \sqrt{r^2+a^2} \sin \theta \cos \phi \left[ \frac{r}{\Delta} \total r + \cot \theta \total \theta - \tan \phi \total \phi \right] \\
&\quad+ \frac{\sin \theta}{\sqrt{r^2+a^2}} ( r \cos \phi - a \sin \phi ) \, U \eqend{,} \raisetag{1.8em}
\end{split} \\
\begin{split}
\tilde{e}^3 &= \sqrt{r^2+a^2} \sin \theta \sin \phi \left[ \frac{r}{\Delta} \total r + \cot \theta \total \theta + \cot \phi \total \phi \right] \\
&\quad+ \frac{\sin \theta}{\sqrt{r^2+a^2}} ( r \sin \phi + a \cos \phi ) \, U \eqend{,} \raisetag{1.8em}
\end{split}
\end{equations}
where we defined
\begin{equation}
U \equiv \frac{M r}{\Sigma} \total t - \frac{M r}{\Delta} \total r + \frac{M r}{\Sigma} a \sin^2 \theta \total \phi \eqend{.}
\end{equation}

The tetrad used by Refs.~\cite{pereiraetal2001}, also in Boyer--Lindquist coordinates, has the form (in our notation)
\begin{equations}[pure_tetrad_pereiraetal]
h^0 &= \sqrt{ 1 - \frac{2 M r}{\Sigma} } \total t + \frac{2 M r a \sin^2 \theta}{\sqrt{ \Sigma ( \Sigma - 2 M r ) }} \total \phi \eqend{,} \\
h^1 &= \sqrt{\frac{\Sigma}{\Delta}} \cos \theta \total r - \sqrt{\Sigma} \sin \theta \total \theta \eqend{,} \\
\begin{split}
h^2 &= \sqrt{\frac{\Sigma}{\Delta}} \sin \theta \cos \phi \total r + \sqrt{\Sigma} \cos \theta \cos \phi \total \theta \\
&\quad- S \sin \theta \sin \phi \total \phi \eqend{,}
\end{split} \\
\begin{split}
h^3 &= \sqrt{\frac{\Sigma}{\Delta}} \sin \theta \sin \phi \total r + \sqrt{\Sigma} \cos \theta \sin \phi \total \theta \\
&\quad+ S \sin \theta \cos \phi \total \phi
\end{split}
\end{equations}
with
\begin{equation}
S^2 \equiv r^2 + a^2 + \frac{2 M r}{\Sigma - 2 M r} a^2 \sin^2 \theta \eqend{.}
\end{equation}
Comparing with Eqs.~\eqref{pure_tetrad_boyerlindquist}, we see that on one hand, the expressions are more complicated because of the square roots, but on the other hand, only the $h^0$ tetrad has a time component $\total t$. Therefore, both tetrads are comparable in their complexity for $M > 0$. However, for the flat background which is obtained for $M \to 0$, the tetrad~\eqref{pure_tetrad_boyerlindquist} is simpler than~\eqref{pure_tetrad_pereiraetal}; they agree only asymptotically for $r \to \infty$ or in the static case $a = 0$. Similar comments apply to the tetrads used by Refs.~\cite{malufetal2002,malufetal2005}, which are of a similar but more complicated form than~\eqref{pure_tetrad_pereiraetal}.

Nevertheless, since both tetrads give the same metric~\eqref{kerrmetric_boyerlindquist}, there exists another Lorentz transformation that connects them: $h^a = \hat{\Lambda}^a{}_b \, \tilde{e}^b$. Since the full expression is very long, we content ourselves again with next-to-leading order for large $r$, where
\begin{splitequation}
\label{hatlambda_asymp}
\hat{\Lambda}^a{}_b &= \delta^a_b - \frac{M}{r} \begin{pmatrix} 0 & \cos \theta & \sin \theta \cos \phi & \sin \theta \sin \phi \\ \cos \theta & 0 & 0 & 0 \\ \sin \theta \cos \phi & 0 & 0 & 0 \\ \sin \theta \sin \phi & 0 & 0 & 0 \end{pmatrix} \eqend{,}
\end{splitequation}
up to terms of order $\bigo{r^{-2}}$.

Since in the first transformation with parameter $\Lambda^a{}_b$ the spin connection was taken to transform inhomogeneously~\eqref{pure_tetrad_spinconn}, the torsion transforms homogeneously according to $\tilde{T}^a = \Lambda^a{}_b T^b$. In the second transformation with parameter $\hat{\Lambda}^a{}_b$, we keep the (now vanishing) spin connection fixed, and hence have again an inhomogeneous transformation~\eqref{torsion_lorentztrafo} of the torsion: $\hat{T}^a = \hat{\Lambda}^a{}_b \, \tilde{T}^b + \total \hat{\Lambda}^a{}_b \wedge \tilde{e}^b$. Since the expression for the transformed torsion $\hat{T}^a$ is even longer than the one for the Lorentz transformation~\eqref{hatlambda_asymp}, we also only give the leading-order terms:
\begin{equations}[pure_tetrad_torsion_asymp]
\hat{T}^0 &= - \frac{M}{r^2} \bar{e}^0 \wedge \bar{e}^1 + \bigo{r^{-3}} \eqend{,} \\
\hat{T}^1 &= \frac{M}{r^2} \sin \theta \, \bar{e}^1 \wedge \bar{e}^2 + \bigo{r^{-3}} \eqend{,} \\
\hat{T}^2 &= - \frac{M}{r^2} \bar{e}^1 \wedge \left( \cos \theta \cos \phi \, \bar{e}^2 - \sin \phi \, \bar{e}^3 \right) + \bigo{r^{-3}} \eqend{,} \\
\hat{T}^3 &= - \frac{M}{r^2} \bar{e}^1 \wedge \left( \cos \theta \sin \phi \, \bar{e}^2 + \cos \phi \, \bar{e}^3 \right) + \bigo{r^{-3}} \eqend{,}
\end{equations}
which agree with~\cite{pereiraetal2001} (after converting to our conventions). However, we note the following issue (which happens also in~\cite{pereiraetal2001,malufetal2002,malufetal2005}): even for $M = 0$, the torsion does not vanish, and instead we have
\begin{equations}
\hat{T}^0 &= \bigo{M} \eqend{,} \\
\hat{T}^1 &= \frac{a^2 \sin(3\theta)}{2 r^3} \bar{e}^1 \wedge \bar{e}^2 + \bigo{M} + \bigo{r^{-4}} \eqend{,} \\
\begin{split}
\hat{T}^2 &= - \frac{a^2}{2 r^3} \Bigl[ \cos(3\theta) \cos \phi \, \bar{e}^1 \wedge \bar{e}^2 \\
&- \cos^2 \theta \sin \phi \left( \bar{e}^1 - \tan \theta \, \bar{e}^2 \right) \wedge \bar{e}^3 \Bigr] + \bigo{M} + \bigo{r^{-4}} \eqend{,} \raisetag{3.8em}
\end{split} \\
\begin{split}
\hat{T}^3 &= - \frac{a^2}{2 r^3} \Bigl[ \cos(3\theta) \sin \phi \, \bar{e}^1 \wedge \bar{e}^2 \\
&+ \cos^2 \theta \cos \phi \left( \bar{e}^1 - \tan \theta \, \bar{e}^2 \right) \wedge \bar{e}^3 \Bigr] + \bigo{M} + \bigo{r^{-4}} \eqend{.} \raisetag{3.8em}
\end{split}
\end{equations}
That is, the limit $M \to 0$ does not actually yield flat space in the pure tetrad approach with the above tetrad, and one has to take in addition $a = 0$ to obtain a vanishing torsion and thus flat space. In contrast, the background tetrad~\eqref{background_tetrad} in the covariant approach including the background spin connection~\eqref{background_spin_conn_form} gives a vanishing background torsion $\bar{T}_{\mu\nu}{}^a$ for all $a$ by construction.

Lastly, we consider the rotating black hole with a coupling to electromagnetism as studied in Refs.~\cite{capozzielloetal2012,ahmedetal2016,capozziellonashed2019}. Unfortunately, in this case it seems that in the pure tetrad approach it is not possible to obtain a black hole with spherical horizon topology, even a static one~\cite[Sec.~4.4]{capozzielloetal2012}, and only flat transverse sections are possible. Since the Kerr--Schild solution presented here has spherical topology, it is thus essentially different from the solutions studied in Refs.~\cite{capozzielloetal2012,ahmedetal2016,capozziellonashed2019}.

\section{Conclusions}
\label{sec_conclusion}

We have shown that the famous Kerr--Schild ansatz for the metric in Einstein gravity that effectively linearises the Einstein equations is also useful in teleparallel gravity, and in the same way linearises the equations of motion for the torsion. In this way, given a background geometry that solves the equations of motion, one finds a new solution by solving a second-order differential equation for a geodesic null vector. While this equation is quite complicated in general, it becomes relatively simple if the background geometry is flat. An important ingredient in teleparallel gravity is the spin connection, which in contrast to Einstein gravity only describes inertial effects coming from a non-inertial frame. Assuming that the character of the frame does not change in the  Kerr--Schild construction, the spin connection is thus fixed by the background geometry, and for a flat background can be determined from the background tetrad solving the zero-torsion condition. Since the torsion scalar $\mathcal{T}$ vanishes in that case, we have furthermore shown that for a flat background geometry any Kerr--Schild solution that is determined in this way is also a solution of $f(\mathcal{T})$ gravity with the same tetrad and spin connection.

We have then exemplified the result by deriving the full torsion for the rotating black hole solution, based on the Kerr--Schild form of the Kerr solution in Einstein gravity. However, the tetrad that appears naturally in the Kerr--Schild form differs from the one used in previous treatments of the Kerr black hole in teleparallel gravity~\cite{pereiravergaszhang2001,lucasetal2009,krssakpereira2015}, and thus the solution for the torsion also differs. Nevertheless, we have shown that one can transform it into the form given previously by a suitable Lorentz transformation, which then also results in the same conserved charges. Since the spin connection was fixed from the beginning, the Lorentz transformation results in an inhomogeneous transformation of the torsion, which is an example of the difficulties pointed out in~\cite{malufetal2018} if the spin connection is determined from the background. On the other hand, the fixed spin connection allows us to obtain solutions of $f(\mathcal{T})$ gravity ``for free'', which are almost certainly \emph{not} anymore solutions for the Lorentz-transformed tetrad and torsion. Our result coincides with the analysis of~\cite{bejaranoetal2015}, who found a rotating black hole solution of $f(\mathcal{T})$ gravity with $\mathcal{T} = 0$ by using the null tetrad associated to the Kerr--Schild form of the Kerr metric; we have shown that this remains true for any Kerr--Schild metric. We have also shown how to transform the solution in the pure-tetrad approach where the spin connection vanishes.

In continuation, it would interesting to see how the general rotating black hole in higher dimensions, and with a de~Sitter or anti-de Sitter background~\cite{gibbonsetal2004} looks like in teleparallel gravity and its $f(\mathcal{T})$ generalisation. For a flat background, the equation for the null vector as well as the geodesic equation are the same as in Einstein gravity, such that one immediately obtains the corresponding solutions in teleparallel gravity and only needs to compute the torsion. However, for a non-trivial background it first needs to be checked if the Kerr--Schild null vector can be taken over identically, or if it needs to be modified. This is particularly important because it has been shown that even the (anti-)de~Sitter background which is a maximally symmetric solution of Einstein gravity, is not maximally symmetric in teleparallel gravity~\cite{coleyetal2019}, and one needs to find a background tetrad and spin connection such that null vector is geodesic. While in the teleparallel equivalent of Einstein gravity, the field equations do not determine the spin connection (and one can thus use the approach of~\cite{krssaketal2018} to fix it by demanding that in the absence of gravity flat space is a solution), this is no longer true for $f(\mathcal{T})$ gravity and other generalisations. It appears that solving for a suitable spin connection is a non-trivial problem in that case in general, even if one assumes a highly symmetric geometry~\cite{hohmannetal2019,jaervetal2019,bahamondeetal2020}. In the Kerr--Schild ansatz we have presented, this difficulty is fortunately absent for a flat background spacetime, but it needs to be checked whether that remains true in the case where also the background is curved.

Furthermore, it would be interesting to also generalise the Kerr--Schild ansatz to symmetric teleparallel gravity \cite{nesteryo1998,beltranetal2019}, where both curvature and torsion vanish and gravitation is described by non-metricity, and to find mappings between those different descriptions. Another venue of generalisation would include more general actions that not only involve the torsion scalar, but also matter fields such as in~\cite{bahamondeetal2020} where a scalar field with quite arbitrary kinetic term and potential was studied. It should be possible to find solutions at least in simple cases, for example including electromagnetic fields where the solution is known in Einstein gravity~\cite{debneyetal1969}; in the general case, the results of~\cite{babichevetal2021} might be helpful. Once the solutions are obtained, one could then study physical implications, for example accretion flows of fluid matter as done in Refs.~\cite{ahmedetal2016,capozziellonashed2019}.

\begin{acknowledgement}
This work has been funded by the Deutsche Forschungsgemeinschaft (DFG, German Research Foundation) --- project nos. 415803368 and 406116891 within the Research Training Group RTG 2522/1. It is a pleasure to thank Martin Kr{\v s}{\v s}{\'a}k and the anonymous referee for comments and references.
\end{acknowledgement}

\appendix

\bibliography{literature}

\end{document}